# Nanoscale Cathodoluminescence Spectroscopy Probing the Nitride Quantum Wells in an Electron Microcope

Zhetong Liu[1,2], Bingyao Liu[1,2], Dongdong Liang[3,4], Xiaomei Li[1], Xiaomin Li[5], Li Chen[1], Rui Zhu[1]*, Jun Xu[1], Tongbo Wei[3,4]*, Xuedong Bai[5]*, Peng Gao[1,2,6]*.

[1] *Electron Microscopy Laboratory, School of Physics, Peking University, Beijing 100871, China.*

[2] *Academy for Advanced Interdisciplinary Studies, Peking University, Beijing 100871, China.*

[3] *Research and Development Center for Semiconductor Lighting Technology, Institute of Semiconductors, Chinese Academy of Sciences, Beijing 100083, China.*

[4] *Center of Materials Science and Optoelectronics Engineering, University of Chinese Academy of Sciences, Beijing 100049, China.*

[5] *Beijing National Laboratory for Condensed Matter Physics, Institute of Physics, Chinese Academy of Sciences, Beijing 100190, China.*

[6] *International Center for Quantum Materials, Peking University, Beijing 100871, China.*

*Corresponding author. Email: tbwei@semi.ac.cn; xdbai@iphy.ac.cn; zhurui@pku.edu.cn; pgao@pku.edu.cn.

## Abstract

To gain a deeper understanding of the luminescence of multiquantum wells and the factors affecting it on a microscopic level, cathodoluminescence combined with scanning transmission electron microscopy and spectroscopy was used to reveal the luminescence of $In_{0.15}Ga_{0.85}N$ five-period multiquantum wells. The composition-wave-energy relationship was established in combination with energy-dispersive X-ray spectroscopy , and the bandgaps of $In_{0.15}Ga_{0.85}N$ and GaN in multiple quantum wells were extracted by electron energy loss spectroscopy to understand the features of cathodoluminescence luminescence spectra. The luminescence differences between different periods of multiquantum wells and the effects on the luminescence of multiple quantum wells owing to defects such as composition fluctuation and dislocations were revealed. Our study establishing the direct correspondence between the atomic structure of $In_xGa_{1-x}N$ multiquantum wells and photoelectric properties, provides useful information for nitride applications.

**Keywords**: Nitride multiquantum wells, defect, cathodoluminescence, scanning transmission electron microscopy.

# 1. Introduction

III-Nitrides are widely used in optoelectronic devices because of their large bandgaps, high thermal conductivities, high electron mobilities, and high breakdown field strengths[1-3]. Among them, the light-emitting diodes (LEDs) prepared using multiple quantum wells (MQWs) are environmentally friendly, have a higher energy efficiency than traditional lighting[4-5], and are widely used in lighting devices[6]. InGaN-based blue LEDs utilize GaN quantum barriers (QBs) between quantum wells (QWs) to limit the carriers, vastly increasing the light efficency[10]. However, their performance is still limited by problems such as the quantum efficiency decrease[7-9], whose mechanism is still unclear owing to the lack of microscopic luminescence characterization.

In recent years, a more in-depth research of the luminescence mechanisms of MQWs has brought luminescence studies to the nanometer scale[11-15], all of which have been enabled by the development of scanning transmission electron microscopy (STEM)-cathodoluminescence (CL). CL is the ultraviolet, visible, and near-infrared light emitted by materials under the excitation caused by the incidence of high-energy electron beams. The luminescence mechanisms can be divided into incoherent and coherent emissions. The material system that we studied was a direct-bandgap semiconductor in which incoherent emission predominates. In terms of the yield of incoherent emissions, the smaller the voltage, the larger the volume of interactions between the electrons and the material, and the better the yield. Therefore, an 80 kV voltage was adopted to ensure not only a high output but also a high spatial resolution. On the one hand, compared with other spectral characterization techniques, such as laser-excited photoluminescence (PL)[16,17], CL has several advantages. For example, the source in CL is super-continuous and has a large energy range; hence, CL is suitable for multiple transitions, including broad bandgap transitions and core-level transitions[18]. On the other hand, compared to SEM-CL[19,20], STEM-CL provides higher spatial resolution and can be combined with high-angle annular dark field (HAADF) image and electron energy loss spectrum (EELS), which can study the optical properties of materials, such as local bandgap changes and defects with nanometer resolution. Therefore, STEM-CL is a powerful method for studying the optoelectrical properties of nanostructures with quantum-size effects in semiconductor optoelectronic devices such as LEDs.

In this study, we performed STEM-CL to characterize the luminescence of five-period $In_xGa_{1-x}N$/GaN MQWs. Three CL luminescence peaks, including yellow band luminescence (YL), GaN, and MQWs were observed in the CL spectra; however, the wavelength and intensity of the luminescence could be affected by many universal defects at the nanoscale. First, the piezoelectric polarization fields caused by the strain in different periods of the quantum well led to a change in the luminescence wavelength. The differences in electron and hole mobility can also influence the luminescence intensity. Second, the intensity of the CL luminescence was weakened or even annihilated owing to the difficulty of electron-hole pair recombination in the region with less In content, and the luminescence peak blue-shifted with decreasing In content. Finally, owing to the release of strain at the dislocation, the nonradiative recombination rate increased, resulting in a decrease in the luminescence intensity. Our study clarifies the microscopic mechanics for failure analysis of MQW-based devices, creating new tunable factors for the design of new optoelectrical devices.

## 2. Results and discussion

The five-period $In_xGa_{1-x}N$/GaN MQWs are grown by the metal-organic chemical vapor deposition (MOCVD) method on graphene (Gr)/$SiO_2$/Si(100) substrates, whose structure is shown in **Figure 1a**. Our low-temperature STEM-CL system is shown in **Figure 1b**, in which the temperature can be reduced to 102 K. The two parabolic mirrors and low-temperature environment significantly improved the signal-to-noise ratio (SNR), enabling better CL collection efficiency. Furthermore, the transmission electrons enabled energy-dispersive X-ray spectroscopy (EDS) and EELS measurements in the same area, as shown in Figure 1c. The atomic content of In and Ga in the wells was found to be 0.15 and 0.85 by EDS, respectively. According to Vegard's Law[21], the band gap of $In_{0.15}Ga_{0.85}N$ was estimated to be 2.75 eV, and the LED emission wavelength was approximately 450 nm. The EELS mapping to characterize bandgaps of $In_xGa_{1-x}N$ MQWs region is shown in **Figure 1c,** and the typical EELS spectra integrated from the QW and QB are shown in **Figure 1d**, where the bandgaps were extracted by a linear fitting method. The bandgap at InGaN QW is ~2.8 eV (~443 nm) and the bandgap at GaN QB is ~3.1 eV (~400 nm). The STEM-CL spectrum of $In_xGa_{1-x}N$ MQWs is shown in **Figure 1e**, which exhibits three distinct peaks. The band diagrams of these peaks are shown in **Figure 2a**. Combined with the EELS characterization, the first peak at ~400 nm originates from the near band edge (NBE) emission of the GaN QBs. The bandgap of the GaN QBs is not the same as that of bulk GaN ($E_{g,GaN}$ = 3.4 eV) because the GaN QBs contain a small amount of In and the effect caused by Cerenkov loss. The peak at ~450 nm is consistent with the front bandgap estimation, which is the NBE luminescence peak of the $In_{0.15}Ga_{0.85}N$ QWs. The last peak at ~580 nm originates from the YL owing to the defect state. [22]

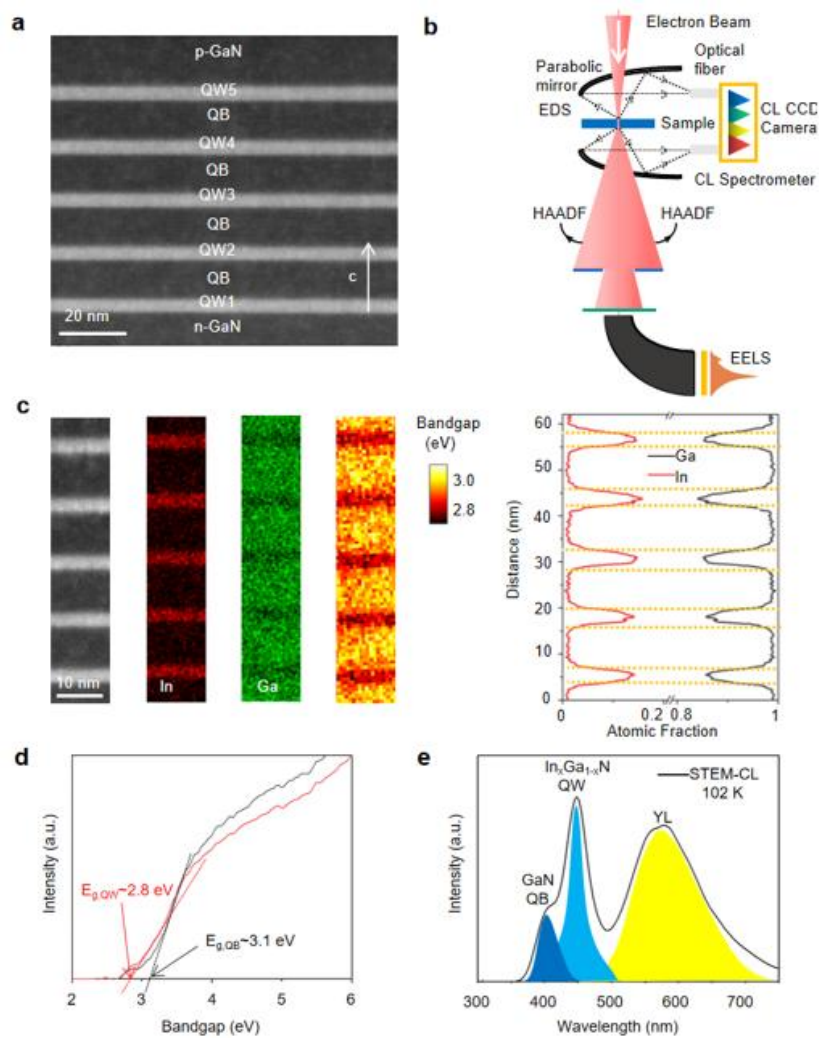

**Figure 1.** STEM-CL characterization combined with the HAADF, EDS, and EELS of $In_xGa_{1-x}N$ MQWs. **a)** HAADF image of $In_xGa_{1-x}N$ LED. **b)** Schematic diagram of STEM-CL,

HAADF, EDS and EELS. **c)** HAADF image, EDS mappings of In (red), Ga (Green), EELS bandgap mapping and atomic fractions of In and Ga of the $In_xGa_{1-x}N$ MQWs. **d)** EELS spectrum of QB and QW. **e)** STEM-CL spectrum of the $In_xGa_{1-x}N$ MQWs.

To further study the luminescence of each period QW at the microscopic level, we conducted line-scanning STEM-CL experiments with a step length of 1 nm and a probe size of ~0.2 nm (**Figure 2b**). The CL spectral lines of each QW were superimposed in terms of the luminescence wavelength, as shown in **Figure 2c**. The NBE luminescence peak of $In_xGa_{1-x}N$ first exhibited a blue shift, followed by a red shift of ~20 nm. The traditional Crosslight APSYS simulation failed to explain this owing to the lack of nonuniform trains in actual MQWs. As shown in **Figure 2d**, owing to the spontaneous polarization of bulk GaN and piezoelectrically induced polarization from the mixing of In element, the combination of the built-in electric field causes energy band bending and electron and hole wave-function separation; this results in the reduction of recombination efficiency and a red shift of the emission wavelength called the quantum-confined Stark effect (QCSE) [23-25]. In the structure of the MQWs, the strains in the periods of the MQWs are not the same. As **Figure 2e** shows, the strains in different periods were obtained by geometric phase analysis (GPA), which showed that the longitudinal tensile strain decreased in the first two periods and then increased as the QW approached p-GaN. This indicates that the piezoelectrically induced polarization electric field also decreases and increases, resulting in the blue/red shift of the luminescence peak[26,27], which subsequently leads to the NBE luminescence wavelength undergoing a blue shift followed by a red shift across different periods.

We also observed that the 4th QW exhibited the strongest light-emitting tendency (**Figure 2f**). Because the electron mobility was fast and the hole mobility was slow[28]. The electrons and holes recombined mostly in the QW near the p-type GaN, where the luminescence is stronger. However, the influence of the growth quality after doping also decreased the luminescence intensity of the last QW, resulting in the final tendency.

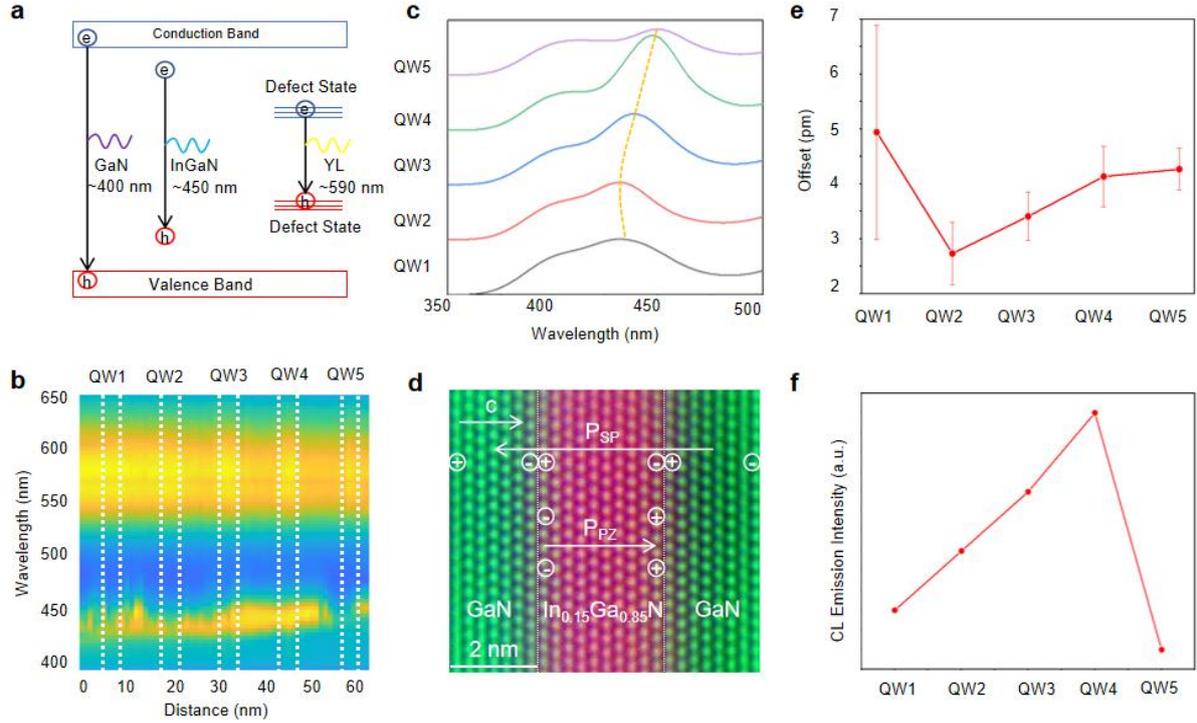

**Figure 2.** CL emission difference of five periods of MQWs. **a)** Band diagram of three CL emission peaks. **b)** STEM-CL mapping across different five periods of MQWs. **c)** The shift of NBE wavelength in QWs. **d)** Schematic diagram of built-in electric field in MQWs. **e)** Spectrum of the strain processing by GPA in QW1-5. **f)** Spectrum of CL emission intensity in QW1-5.

It is worth noting that there is usually a non-uniform composition in the MQWs. A typical QW with partially missing In is shown in **Figure 3a-b**, and the atomic fraction change of the In content is shown in **Figure 3c**. We performed a STEM-CL characterization in this region (**Figure 3d**) and discovered that the NBE emission peak of the QW decreased with the composition of In (**Figure 3e**). Moreover, the CL luminescence intensity weakened or even disappeared in the In-deficient region (**Figure 3f**). This can be explained from the perspective of recombination. According to the different methods of energy release, recombination can be divided into radiation and non-radiation recombination. Direct radiation recombination, which is the primary form of radiation recombination, refers to the process of direct recombination between conduction-band electrons and valence-band holes. It can also be carried out through the recombination center to release energy in other ways, which is called non-radiative recombination. Because the CL can only collect signals of radiation recombination, the reduction of In content makes it difficult to achieve the electron-hole pair recombination, thus enhancing the non-radiative recombination, which weakens the intensity of the CL luminescence.

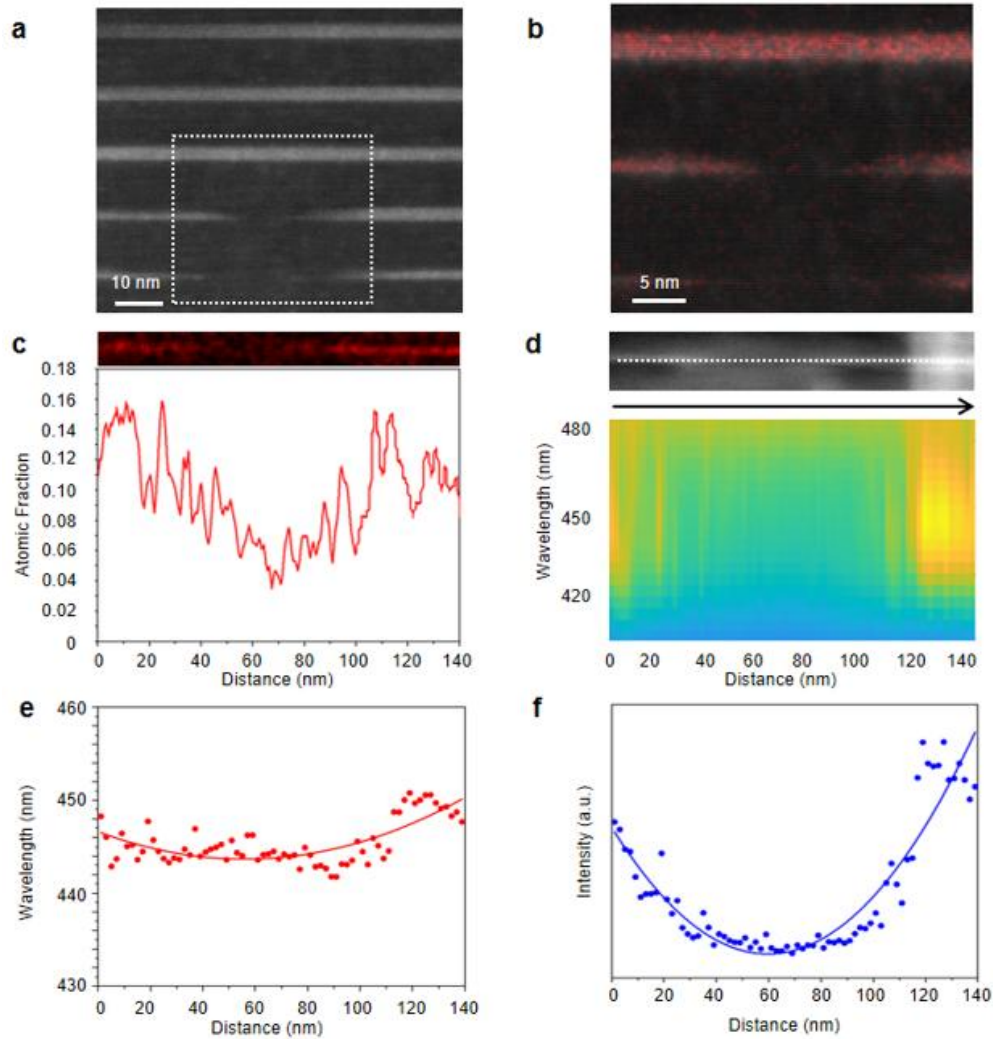

**Figure 3.** Effect of In composition fluctuation on CL emission. **a,b)** HAADF image and EDS mapping of local In composition fluctuation in MQWs. **c)** Atomic fraction of In element in QW1. **d)** STEM-CL mapping across In composition fluctuation region in QW1. **e,f)** Spectra of wavelength and intensity of NBE across In composition fluctuation region in QW1.

Additionally, various defects in GaN can affect the luminescence of the MQWs. A HAADF image of a threading dislocation (TD) in the MQWs is shown in **Figure 4a**, which is judged to be a mixed-type dislocation. Previous studies have found it hard to explain the structure-activity relationship between the dislocation and luminescence of MQWs owing to the limitation of spatial resolution.[29-32]

As shown in **Figure 4b**, we studied the luminescence of the mixed dislocation region in the MQWs. The NBE luminescence peak of $In_xGa_{1-x}N$ QW around the dislocation has a blue shift of ~15 nm (**Figure 4c**). As **Figure 4d** shows, the strain is released at the dislocation, and the strain-induced polarization electric field is depressed, which leads to a blue shift of the luminescence peak near the dislocation. On the other hand, **Figure 4e** shows that the luminescence intensity at the mixed dislocations was significantly weakened, which can be attributed to recombination. As shown in **Figure 4f**, the mixed-type dislocation in n-type GaN is an electron trap with a negative charge, which has a scattering effect on the carriers. The energy generated by the electron and hole recombination is likely to be captured by another electron, resulting in Auger recombination and the formation of a nonradiative recombination center[33-35]. The nonradiative recombination rate thus increases near the dislocation, leading to a decrease in minority carriers and luminescence intensity.

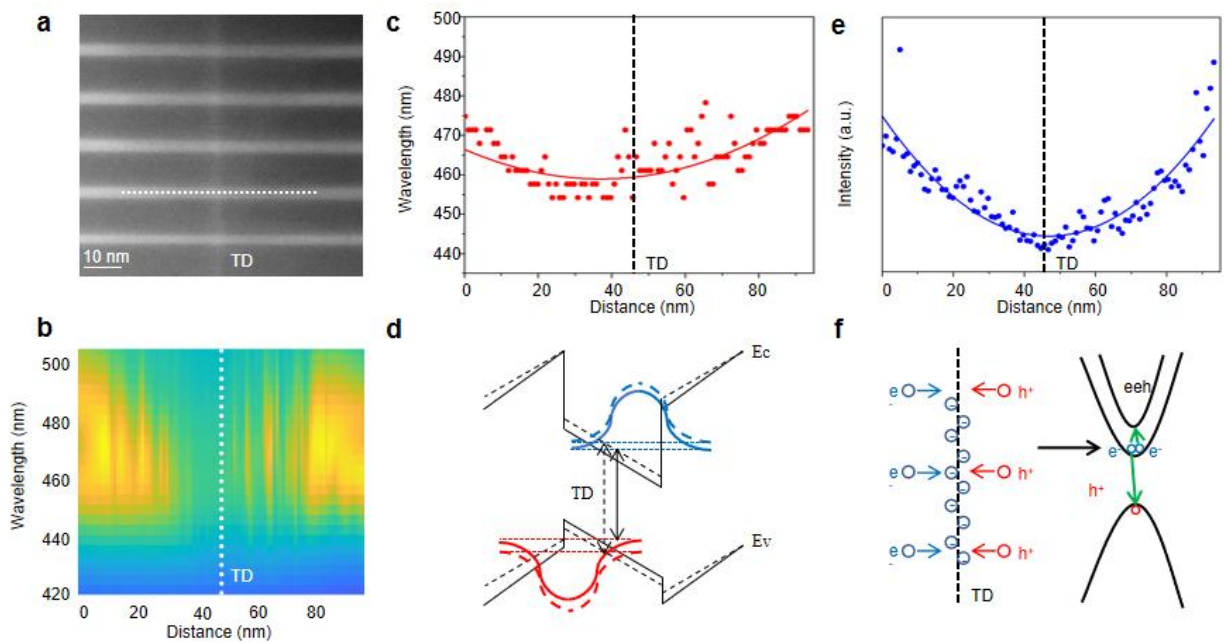

**Figure 4.** Effect of the TD on CL emission. **a, b)** HAADF image and STEM-CL mapping of TD in MQWs. **c)** Spectrum of wavelength of NBE across the TD. **d)** Schematic diagram of the effect of the TD on the built-in electric field. **e)** Spectrum of the intensity of NBE across the TD. **f)** Schematic diagram of the non-radiative recombination mechanism around the TD.

## 3. Conclusions

Using STEM-CL characterizations at 102 K, we studied the luminescence behavior of five periods of $In_{0.15}Ga_{0.85}N/GaN$ MQWs and the influence of defects (including component undulations and dislocations) on luminescence. We found that the strain and defects of the material would affect the recombination process and piezoelectrically-induced polarization electric field, leading to changes in the wavelength and intensity of the MQWs luminescence. Thus, the direct correspondence between the atomic structure of $In_xGa_{1-x}N$ MQWs and photoelectric properties was established. These microscopic changes affect the macroscopic performance of the device, so they should be considered during the fabrication of MQWs and the designing of the device.

## 4. Methods

**MOCVD process of the GaN and blue LED on $Gr/SiO_2/Si(100)$ substrate:**
Trimethylgallium (TMGa), trimethylaluminum (TMAl), and $NH_3$ were used as Ga, Al, and N precursors for growing AlN and GaN films; The III-nitride films were grown on the $Gr/SiO_2/Si(100)$ substrate using the Veeco K300 MOCVD chamber. First, the high temperature (HT)-AlN was grown at 1200 °C for 6 min with the $NH_3$ flow of 1000 sccm and TMAl flow of 50 sccm, respectively. Then GaN growth procedure: first, the 1st-GaN layer was grown at 1050 °C for 40 min. Then 5 periods of $In_xGa_{1-x}N/GaN$ MQWs layer were grown at 735 °C/834 °C with 3 nm InGaN well layers and 12 nm GaN barriers. The active layers were capped with a p-GaN layer deposited at 950 °C with the bis-cyclopentadienyl magnesium (Cp2Mg) flow of 120 sccm, followed by an annealing process at 720 °C for 10 min under $N_2$ ambient.

**Electron microscopy characterizations and analysis:**
The cross-sectional TEM specimen was made by the ThermoFisher Helios G4 UX focused ion beam system. The HAADF and EDS mapping were performed using FEI Titan Cubed Themis G2 300 spherical aberration corrected STEM, operated at 300 kV. The convergence semi angle was 30 mrad and the collection semi angles of HAADF was 39 to 200 mrad. The camera length in HAADF mode was set as 145 mm. The STEM-CL spectra were taken on JEOL Grand ARM 300 equipped with a Vulcan CL detector, operated at 80 kV.

## Acknowledgments

The work is supported by the National Key R&D Program of China (2019YFA0708202), the National Natural Science Foundation of China (11974023, 52021006, 61974140, 12074369), the "2011 Program" from the Peking-Tsinghua-IOP Collaborative Innovation Center of Quantum Matter, and the Youth Supporting Program of Institute of Semiconductors. We acknowledge Electron Microscopy Laboratory of Peking University and Institute of Physics of Chinese Academy of Sciences for the use of electron microscopes.

## Author Contributions

P.G. and Z.T.L. conceived the idea and designed this work. Z.T.L. performed the transmission electron microscopy and cathodoluminescence experiments under the direction of P.G., X.D.B., X.M.L., X.M.L., J.X., L.C. and R.Z.. D.D.L. designed and performed the growth experiments under the direction of T.B.W.. Z.T.L., B.Y.L. and P.G. co-wrote the manuscript. All authors discussed the results and comment on the manuscript.